\documentclass[useAMS,usenatbib,letterpaper]{mn2e}
\usepackage[dvips]{graphicx}
\usepackage{latexsym}
\usepackage{amsmath}
\usepackage{amsfonts}
\usepackage{amssymb}
\usepackage{multicol}
\usepackage{array}

\def\lea{\mathrel{<\kern-1.0em\lower0.9ex\hbox{$\sim$}}}
\def\gea{\mathrel{>\kern-1.0em\lower0.9ex\hbox{$\sim$}}}
\def\leq{\mathrel{<\kern-1.0em\lower0.9ex\hbox{$-$}}}
\def\geq{\mathrel{>\kern-1.0em\lower0.9ex\hbox{$-$}}}

\title[I.  Velocity Dispersion Measurements of
  Central Coma Galaxies]{Kinematic Properties and Stellar Populations of Faint
  Early-Type Galaxies.  I.  Velocity Dispersion Measurements of
  Central Coma Galaxies}
\author[Matkovi\'c \& Guzm\'an]{A. Matkovi\'c$^{1}$\thanks{E-mail:
matkovic@astro.ufl.edu (AM)} and R. Guzm\'an$^{1}$\\
$^{1}$Department of Astronomy, University of Florida, P.O.\
Box 112055, Gainesville, FL 32611, USA\\
}
\begin{document}

\date{Accepted 2005 June 16. Received 2004 November 25}

\pagerange{\pageref{firstpage}--\pageref{lastpage}} \pubyear{2004}

\maketitle

\label{firstpage}

\begin{abstract}

We present velocity dispersion measurements for 69 faint early--type
galaxies in the core of the Coma cluster, spanning $-22.0 \lea M_R
\lea -17.5$ mag.  We examine the $L - \sigma$ relation for our sample
and compare it to that of bright ellipticals from the literature.  The
distribution of the the faint early--type galaxies in the $L-\sigma$
plane follows the relation $L \propto \sigma^{2.01\pm0.36}$, which is
significantly shallower from $L \propto \sigma^{4}$ as defined for the
bright ellipticals.  While increased rotational support for fainter
early--type galaxies could account for some of the difference in
slope, we show that it cannot explain it.  We also investigate the
Colour--$\sigma$ relation for our Coma galaxies.  Using the scatter
in this relation, we constrain the range of galaxy ages as a function
of their formation epoch for different formation scenarios. Assuming a
strong coordination in the formation epoch of faint early--type
systems in Coma, we find that most had to be formed at least 6 Gyrs
ago and over a short 1 Gyr period.

\end{abstract}


\section{Introduction}

Dwarf elliptical galaxies (dEs) are low luminosity spheroidal systems
with $M_B > -18$ mag (Sandage \& Binggeli 1984) that have a low
surface brightnesses, $\mu_{e,V}>22$ mag arcsec$^{-2}$ (Ferguson \&
Binggeli 1994).  Over the past two decades there has been considerable
interest in studying dEs, despite the difficulty associated with
observing these objects.  In particular, there has been a number of
extensive photometric studies that have concentrated on comparing dEs
to the bright elliptical galaxies with $M_B \leq -20.50$ mag (Graham
\& Guzm\'an 2003).  Two main arguments support the hypothesis that
they are structurally distinct classes.

Firstly, many authors in the past argued that de Vaucouleur's
$R^{1/4}$ law (de Vaucouleurs 1948) best fits the light profiles of
bright Es.  Meanwhile, Faber \& Lyn (1983) and Binggeli, Sandage \&
Tarenghi (1984) showed that an exponential profile describes dwarf
ellipticals.  Secondly, as Kormendy (1985) notes, dEs and bright Es
fall almost perpendicular to each other in the effective surface
brightness--luminosity plot, $\mu_e-L$ and are clearly different in
the luminosity--central surface brightness, $L-\mu_0$, plot.
Furthermore, dEs and bright Es follow a different $\mu_e-R_e$ relation
(e.g. Wirth \& Gallagher 1984; Capaccioli, Caon \& C'Onforio 1992).

More evidence to strengthen the proposed dichotomy exists.  Dwarf
ellipticals seem to lie off the Fundamental Plane, the relation
between the surface brightness at the effective radius, $\mu_e$,
effective half-light radius, $R_e$, and the velocity dispersion,
$\sigma$ (Bender, Burstein, \& Faber 1992; de Carvalho, \& Djorgovski
1992; Peterson, \& Caldwell 1993).  This disparity has been
interpreted as a difference in the formation mechanism for dwarf and
bright Es.

Nonetheless, there are many studies arguing for continuity in the
dwarf-bright family.  Just to mention a few: Caldwell 1983; Caldwell
1987; Caldwell \& Bothun 1987; Ferguson \& Sandage 1988; Hudson
\textit{et al.}~1997; Jerjen \& Binggeli 1997; Jerjen, Binggeli, \&
Freeman 2000; and Karachentsev, \textit{et al.}~1995.  Some of these
studies show that dEs exhibit the same central surface brightness and
absolute magnitude relation as bright Es.  Caldwell (1983) also
pointed out that a continuous trend exists between colour and
luminosity, while Caldwell \& Bothun (1987) show the same continuity
for the luminosity--metallicity relation.

Graham \& Guzm\'an (2003; hereafter GG03)(see also Guzm\'an \textit{et
al.}~2003) offer a possible resolution of the differing views.  They
point out that the dichotomy in the luminosity--effective surface
brightness relation, $M_B-\mu_e$, and the luminosity--effective radius
relation, $M_B-R_e$, is a direct consequence of the linear relations
between the luminosity, the central surface brightness, $\mu_0$, and
the light profile shape, $n$.  Furthermore, they argue that dEs and
intermediate luminosity ellipticals follow a continuous sequence up to
$M_B \sim -20.5$ mag.  At this point bright ellipticals start showing
evidence of evacuated cores, possibly coalescing black holes, causing
their central surface brightness to decrease with increasing
luminosity (GG03; Graham 2004 and references therein).  The most
massive Es may thus be the exception and not the rule to the empirical
correlations defined by early--type galaxies that include the low
luminosity galaxies and range over $\sim 8$ magnitudes (see GG03).

Up until the past couple of years most information on spectroscopic
properties of dEs came from their line strength indices (Held \& Mould
1994; Gorgas \textit{et al.}~1997; Mobasher \textit{et al.}~2002;
Moore \textit{et al.}~2002), and a handful of velocity dispersion
measurements.  Although more difficult to obtain due to low surface
brightness of these objects, the number of papers in the literature
which include velocity dispersion measurements of dEs in different
clusters has increased (Bender \& Nieto 1990; Brodie \& Huchra 1991;
Held \textit{et al.}~1992; Bender, Burstein, \& Faber 1992; Peterson
\& Caldwell 1993; Bernardi \textit{et al.}~1998; Melhert \textit{et
al.}~2002; Hudson \textit{et al.}~2001; De Rijcke \textit{et al.}~
2001; Simien \& Prugniel 2002; Pedraz \textit{et al.}~2002; Moore,
Lucey, Kuntschner, \& Colless 2002 (hereafter MLKC02); Geha,
Guhathakurta, \& van der Marel 2002, 2003; Guzm\'an \textit{et al.}~2003;
Bernardi \textit{et al.}~2003; Smith \textit{et al.}~2004, van Zee,
Skillman, \& Haynes 2004; De Rijcke \textit{et al.}~2004).  The NOAO
Fundamental Plane survey (NFPS) of Smith \textit{et al.}~2004 is the
largest compilation including velocity dispersion measurements in
low-redshift galaxy clusters up to date.  Our Coma sample, although
significantly smaller than that of the NFPS is complementary to this
study.  It includes a statistically representative sample of faint
early-type galaxies in different environments within a cluster.  Such
sample is essential for testing current ideas on the formation and
evolution of dEs.

Via recent studies, dEs have also been linked to the Butcher-Oemler
effect (Butcher, \& Oemler 1978; Butcher, \& Oemler 1984).
Observations of distant clusters reveal the existence of numerous
star-forming, low-mass 'blue disk' galaxies in clusters at $z\sim0.4$.
These galaxies, as shown by HST, are distorted small spirals which
have disappeared from the present day clusters.  The fate of these
galaxies remains one of the most important unanswered questions in
modern cosmology.  The 'galaxy harassment' model Moore, Lake \& Katz
(1996; 1998) explains how the dwarf spiral galaxies in clusters may evolve
into today's population of cluster dEs due to encounters with brighter
galaxies and the cluster tidal field.  The galaxy harassment model
predicts differences in the properties of dEs located in the inner,
high density, and outer, low density, regions of the clusters.

This is the first paper of a series in which we will characterise the
kinematics and stellar populations of dEs and other low luminosity
early--type galaxies as a function of the environment. Given the
difficulty in distinguishing between dEs, dS0s, and dwarf spirals in
Coma from ground-based images, we refer to all these objects as
`early--type' galaxies.  For convenience, we define `faint'
early--type galaxies as those with $M_B > -20.50$ mag, and the
`bright' ellipticals with $M_B \leq -20.50$ mag. Here we describe
spectroscopic observations and velocity dispersion ($\sigma$)
measurements of $69$ faint early--type galaxies in the central
$30\arcmin \times 30\arcmin$ region of the Coma cluster.  We
investigate whether these galaxies follow the luminosity--velocity
dispersion ($L-\sigma$) relation derived for bright ellipticals, and
discuss the constraints on their formation epochs provided by the
colour--$\sigma$ ($C-\sigma$) relation.  In future papers, we will
test the implications of the 'galaxy harassment' scenario by comparing
the internal kinematics and stellar populations of the early--type
galaxies in the core and the outskirts of the Coma cluster.  Section 2
describes the photometric and spectroscopic observations of our sample
galaxies, including the sample selection.  In Section 3, we describe
the data reduction technique and the velocity dispersion measurements.
We investigate the $L-\sigma$ relation in Section 4, and the
$C-\sigma$ relation in Section 5. A summary of our results is provided
in Section 6.

\section[]{Observations}
\subsection{Sample Selection}

The selection of faint early--type galaxies in the Coma cluster was
done using the photometry in $U, B$ and $R$ bands.  We obtained the
images with the WIYN/MiniMo and INT/Wide Field Camera.  To select the
faint early--type galaxy candidates in the central $30\arcmin \times
30\arcmin$ region of the Coma cluster we used the B--R vs. B
colour-magnitude plane.  We applied an absolute luminosity cutoff at
$M_{B} \geq -17.3$ mag (Ferguson, \& Binggeli 1994; corresponding
to apparent $B \gea 17.5$ mag at the distance of the Coma
cluster\footnote{Throughout the paper we use $H_0$ = 70 km s$^{-1}$
Mpc$^{-1}$, and a distance modulus for Coma cluster of 35.078, ($d =
99$ Mpc).}).  To minimize contamination at the faint end by the
background field disk galaxies at $z<0.2$, we applied another cutoff
using the (U--B) vs. (B--R) colour-colour diagram at $0.2 < (U-B) < 0.6$
mag, and $1.3<(B-R)< 1.5$ mag.

\subsection{Spectroscopic Observations}
We observed the Coma cluster faint early--type galaxies during 1998
May 23--26, and 1999 May 14--19 on the 3.5 m WIYN telescope at Kitt Peak
National Observatory with the multi-fibre spectrograph HYDRA.  We used
the 600 l mm$^{-1}$ grating in the $2^{\rm nd}$ order, and the blue
fibre cable, which we chose for its transmission at the desired
wavelengths.  The selected grating allowed us to observe in the
wavelength range of $\Delta\lambda=4120-5600$ \AA, which is optimal
for discerning some of the most prominent absorption features of the
faint early--type galaxies including molecular G-band,
H$\gamma$, H$\beta$, Mg$_2$, and Fe$\lambda 5350$.  With this setup we
achieved the dispersion of $\sim 0.705$ \AA~px$^{-1}$, while our
instrumental resolution of FWHM $=1.91$ \AA~ allowed us to detect
velocity dispersions down to 35 km s$^{-1}$.  This is assuming that we
can measure velocity dispersions up to 30 percent better than the
instrumental resolution for galaxies with SNR $>15$.

The HYDRA multi-fibre spectrograph has $\sim100$ fibres each with
3$\arcsec$ diameter.  Thus, it is well suited for the detection of the
faint early--type galaxies, which typically have a half light radius
of $\sim 2\arcsec$ at the distance of Coma (GG03).  We observed
$\sim45$ galaxies and $\sim45$ adjacent sky spectra for each HYDRA
setup.  The galaxy sample was divided into 3 different groups
depending on the exposure times to achieve SNR $\geq 15$.  We observed
the brightest galaxies $b_j \leq 17.5$ mag for a total integration
time of 4 hours, objects with $17.50 < b_j \leq 18.5$ mag for 8 hours,
and the faintest objects $b_j >18.5$ mag for 16 hours.  In addition,
we obtained spectra of template stars representative of the prevailing
stellar-population of dEs, primarily G and K--type stars.  The sample
also included $\sim30$ bright elliptical galaxies observed in previous
studies in order to assess any systematic effects.  Sample spectra of
4 galaxies with different luminosities and SNR are given in Figure 1.
All the candidates had enough signal for radial velocity measurements
from which we conclude that $100$ percent have the range of recession
velocities of 4,000--10,000 km s$^{-1}$, consistent with membership in
the Coma cluster (Colless \& Dunn 1996).

\section{Data Reduction and Measurements}
\subsection{Basic Data Reduction}

We used the basic and multi-fibre spectral reduction tasks within
IRAF\footnote{Image Reduction and Analysis Facility. Distributed by
the National Optical Astronomy Observatories, which is operated by
AURA (Association of Universities for Research in Astronomy, Inc)
under cooperative agreement with the National Science Foundation.} to
reduce our observations.  The data was first trimmed accordingly and
corrected for the overscan region.  We removed the cosmic rays with
FIGARO within the STARLINK software program.  For the remaining
reduction we used the `dohydra' task within IRAF, which provided:
aperture identification; tracing of the apertures; flat field
correction for the pixel to pixel sensitivity and the different
throughput from fibre to fibre; removal of internal reflections within
the spectrograph (scattered light); the wavelength calibration; and
the sky subtraction.  The largest rms for the wavelength calibration
was 0.02 \AA.

\subsection{Velocity Dispersion and Radial Velocity Measurements}

Velocity dispersions, $\sigma$, were measured from galaxy spectra with
the software REDUCEME (Cardiel 1999).  The program implements the
Fourier quotient method (described by Gonz\'alez-Gonz\'alez 1993),
originally introduced by Sargent \textit{et al.}~(1977), to measure the
velocity dispersions.  The Fourier quotient method assumes that
observed galaxy spectra can be described as a convolution between the
spectral characteristics of the stellar population, the broadening
function, and the effective response function of the instrument.
Using an initial guess of the velocity dispersion, the program
calculates a broadening function described by the dispersion relation
and models a galaxy spectrum as a convolution of the broadening
function and an optimal stellar synthesis spectrum.  This template for
the galaxy is produced by combination of the different star templates.
Via $\chi^2$ minimization between the galaxy and the model spectra,
the best value of the velocity dispersion for the broadening function
is determined.

We tested the stability of the software for a range of values of the
involved parameters.  The only difference in $\sigma$ measurements
occurred when we changed the wavelength range used to calculate the
velocity dispersions and when we altered the tapering fraction.  The
measured velocity dispersion varied by $10$ percent with the choice of
the spectral region, and up to $20$ percent for the faintest galaxies
when varying the tapering fractions.  We conducted a set of tests
altering either the tapering fraction for a specific wavelength range,
or the wavelength range for a specific tapering fraction.  We found
that the most stable solutions occurred for a tapering fraction of 0.25
and when we trimmed the spectra at the edges.  We chose to trim the
spectra by preserving the largest rest frame wavelength range possible:
$4150-5400$ \AA.~

\subsection{Uncertainty Measurements}

We estimated the uncertainties in velocity dispersion measurements by
the `boot--strapping' technique implemented by J. Gorgas (private
communication) within the REDUCEME software.  Assuming that the noise
in the galaxy spectrum is dominated by Poisson noise, the program uses
Monte-Carlo realizations to produce simulated galaxy spectra with
similar SNR to the original galaxy.  For each galaxy we ran 50
simulations, after confirming that the results were the same for 100
simulations.  The program then measures the velocity dispersion of
each simulated galaxy and via $\chi^2$ minimization estimates the
error in the velocity dispersion measurements, which also accounts for
the template mismatch.

We have checked internally and externally for systematical offsets in
the velocity dispersion measurements in our sample of the core faint
early--type galaxies.  The external check consisted of comparing our
measurements with those comparable in SNR and resolution in the
literature, (MLKC02 and NFPS).  Figure 2 shows the comparison of
$\sigma$ measurements and includes the uncertainty in the $\sigma /
\sigma_{Lit}$.  This plot shows no systematic offsets between our and
the two literature samples.  Furthermore, the median of the
comparison, 1.01, and the rms scatter of the points, 0.10, indicate
that our $\sigma$ measurements are in good agreement with those of
both NFPS (open triangles) and MLKC02 (solid triangles).  We also note
that some galaxies seem to be inconsistent with the average value when
taking their uncertainties into account.  This would imply that the
uncertaities in the $\sigma$ measurements are possibly underestimated.
However, it is not possible to determine whether it is the literature,
ours, or both uncertaities that are underestimated.



The internal check was done in the same way as the external, except
that in this case $\sigma$ was measured using only half of their
exposures in our own data.  We found no inconsistencies in the
$\sigma$ measurements nor their error measurements for our sample of
faint early--type galaxies, but the error measurements for our bright
ellipticals were, on average, underestimated by 30 percent.  We have
therefore increased the uncertainties of the bright Es by 30 percent.

In addition, our analysis excludes galaxies with SNR $< 15$, since
below this value the fit of the model spectra to the galaxy was
uncertain.  We also investigate the effects noise has on the velocity
dispersion measurements.  The following method closely resembles the
analysis of J$\o$rgensen, Franx \& Kj$\ae$rgaard (1995; hereafter
JFK95).  We selected a template star that best fits a typical galaxy
with a high SNR, and broadened it by convolving it by Gaussians with
$\sigma$ ranging from 35 to 100 km s$^{-1}$.  Different amounts of
noise were added to each spectra so the SNR would yield values ranging
10--50.  The unaltered spectra of the star was used as a template,
while we measured the velocity dispersion of the broadened and lower
SNR spectra.  The final $\sigma$ was derived after 1000 simulations
using the bootstraping method.  Figure 3 shows the percentage
difference between the 'galaxy' spectra (template spectra broadened to
50 km s$^{-1}$) and the observed $\sigma$, i.e. the same broadened
galaxy with different amount of noise.  All our simulated spectra have
a slightly overestimated measurement of the velocity dispersion.  The
effect is larger for galaxies with a smaller velocity dispersion.  For
example, a galaxy with $\sigma=35$ km s$^{-1}$ and SNR of 15 has
$\sigma$ overestimated by $\sim 6$ percent.  JFK95 find that a galaxy
with $\sigma=65$ km s$^{-1}$ is overestimated by $\sim 4$ percent, and
a galaxy with $\sigma=100$ km s$^{-1}$ by $\sim 1$ percent which is in
good agreement with our measurements.  We chose not to correct for
this systematic effect since it is significantly smaller than the
uncertainties in the $\sigma$ measurements for majority of the
galaxies in our sample.  Implications of this effect on the $L-\sigma$
relation are discussed in section 5.

\section{Results}

The $\sigma$ and radial velocity measurements for 87 early--type
galaxies we observed in the Coma cluster are presented in Table 1.
For the following analysis we only used 72 objects with SNR $\geq 15$.
We classify galaxies with $M_R \leq -22.17$ mag as bright Es.  Our
sample includes 69 galaxies with $M_R \geq -22.17$ mag classified here
as faint early--type galaxies and 3 bright Es.We include other samples
from the literature with $\sigma$ measurements and plot the galaxies
in diagnostic diagrams to characterize their kinematic properties.

In Figure 4 (left) we show the $L-\sigma$ relation including galaxies
with $\sigma$ measurements from MLKC02, Hudson \textit{et al.}~(2001),
EFAR (Colless \textit{et al.}~2001), and our sample.  The total number
of galaxies is 167, where 24 galaxies are classified as bright Es.  In
case of multiple $\sigma$ measurements from the literature, we used a
minimum variance weighted average of velocity dispersions and
calculated the error in the weighted average.  When those galaxies
were in common with our sample, we used the same process but adopted
the symbol used for this paper (triangles) in the figure.  Table 1
lists the $\sigma$ values measured in our sample.


We have derived the photometry in $M_R$ from Guti\'errez \textit{et
al.}~(2004) for the MLKC02 and our sample, while the other catalogues
included their own photometry.  We averaged the actual luminosities for
galaxies with multiple photometry.  The uncertainty in the average of
the magnitudes was 0.015 mag, except for 3 galaxies, which we excluded
from the sample since their uncertainty was large and most likely due
to a catalogue mismatch.  Out of our sample of 72 galaxies with $\sigma$
measurements, 17 were not in the Gutierrez list.  In this case we have
derived $R$ (Johnson filter) by using the $b_j$ magnitudes listed in
the GMP catalogue.  We used a least squares fit to objects that had both
$b_j$ and $R$ and obtained the transformation $R=(1.050\pm0.030)
b_j-2.422\pm0.529$ between these two magnitudes.

The $L - \sigma$ relation in Figure 4 (left) exhibits a curvature or a
change of slope.  To allow for a comparison with earlier studies we
perform least-squares fits to the bright Es and faint early--type
galaxies separately.  We partition the bright Es from the other
early--type galaxies with the dotted line at $M_R \geq -22.17$ mag.
The dashed line represents a most recent $L \propto \sigma^n$ from
the literature (Forbes, \& Ponman 1999), where $\log \sigma =-0.102
M_B + 0.243$, corresponding to $n=3.92$.  We obtained the ordinary
least squares fit (OLS, as described by Feigelson, \& Babu 1992) for
all the galaxies excluding bright Es.  The dash-dot line represents
the fit which minimizes the residuals in $M_R$, the dash-dot-dot
minimization in $\log \sigma$, and the solid line is the bisector
line.  The details of the fits are summarized in Table 2.  Only the
galaxy with a very large observational error from MLKC02 (see Figure
4) is excluded from the linear regression.

Another diagnostic diagram used for the characterization of the low
luminosity early--type galaxies and a comparison of their properties to
those of the bright Es is the $C-\sigma$ relation.  Figure 5 shows the
Johnson B--R mag (from Trenham, unpublished data) vs. $\log \sigma$
for galaxies in our sample, including some galaxies with $\sigma$
measurements from the literature.  In case of the galaxies in common
with the literature, we performed a weighted average on the velocity
dispersions and their errors.  We plot only galaxies that have $B-R$
measurements, except in the case of the Hudson \textit{et al.}~(2001)
sample where we have derived the colours from the GMP catalogue.  The
transformation between the GMP $b-r$ colours and $B-R$ of our sample
was: $B-R=(1.128\pm0.098) (b-r)_j -0.535\pm0.178$.  The derived
$C-\sigma$ relations minimizing residuals in either $\log \sigma$,
B--R, or using the bisector for all the galaxies in the figure are
found in Table 3.  The right-hand panel includes galaxies from
clusters observed by Faber \textit{et al.}~(1989) where we used the
colour transformation $(B-R)=(B-V)+0.71$ from Fukugita, Shimasaku \&
Ichikawa (1995).

\section{Discussion}
\subsection{L-$\sigma$ Relation for Faint Early--type Galaxies}

Faber \& Jackson (1976) showed that luminosity of bright Es correlates
well with velocity dispersion ($\sigma$) for these galaxies.  The
$L-\sigma$, or Faber-Jackson, relation can be expressed as $L \propto
\sigma^n$, where $n$ was originally $\approx 4$ (Faber \& Jackson
1976; Sargent \textit{et al.} 1977; Schechter \& Gunn 1979; Schechter
1980; Tonry \& Davis 1981; Terlevich \textit{et al.}~1981).  Tonry
(1981) was the first to note a slight change of slope in the
$L-\sigma$ relation suggesting that $n \approx 4$ for more luminous
objects while $n \approx 3$ for fainter galaxies.  This result was
confirmed by Davies \textit{et al.}~(1983) who found $n=4.2\pm0.9$ for
galaxies brighter than $M_B=-20$ ($M_R < -21.67$) and $n=2.4\pm0.9$
for those fainter than this magnitude, and Held \textit{et
al.}~(1992) who found $n=2.5$ for dEs.

Unfortunately, the data samples of Tonry (1981), Davies \textit{et
al.}~(1983) and Held \textit{et al.}~(1992) only included a dozen of
the faint early--type galaxies.  To further investigate the $L -
\sigma$ relation for a wide range of luminosities we present a sample
of 143 galaxies with $-22 \lea M_R \lea -17.5$ mag.

The $L-\sigma$ relation (Figure 4) derived for this large sample
exhibits a change of slope; the slope of faint early--type galaxies is
shallower than that of bright ellipticals.  Following the results of
previous studies and including these in our data set we find that the
value of $n \sim 4$ fits the bright E end of the diagram.  In contrast
to bright Es, we obtain $L \propto \sigma^{2.01 \pm 0.36}$ for faint
early--type galaxies (adopting the bisector fit).  This relation spans
a range of 4.5 magnitudes fainter than $M_R= -22.17$ mag, the lower
limit of bright Es, and is the largest sample of the faint early--type
galaxies in a single cluster thus far.  Our result derived for 143
galaxies is consistent with $L \propto \sigma^{2.4 \pm 0.9}$ derived
by Davies \textit{et al.}~(1983) for their 14 faint ellipticals and
that of De Rijcke \textit{et al.}~(2004), and it is inconsistent with
the standard Faber-Jackson relation.  This raises intriguing questions
concerning the physical processes responsible for the change of slope
in the $L-\sigma$ relation.

We note that our galaxies exhibit a small systematic offset depending
on their SNR, as discussed in section 3.2.  However, if corrected for
this effect of $\sim 6$ percent at the lowest $\sigma$ and lowest SNR
galaxies, the slope of the faint early-type galaxies would be even
shallower than derived in this paper. 

A feasible explanation of the different slope between bright Es and
faint early--type galaxies may be due to the presence of other types
of galaxies at lower luminosities.  We used the photometry of
Guti\'errez \textit{et al.}~(2004) to classify galaxies from the
present combined sample.  In cases where the bulge-to-total (B/T)
luminosity ratios from Guti\'errez \textit{et al.}'s data were
unreliable, we used the NASA/IPAC Extragalactic Database (NED).  We
defined galaxies with $B/T=1.0$ as bulge-dominated, $0.5 < B/T < 1.0$
as bulge+single exponential component galaxies, and $B/T<0.5$ as
single exponential component.  This notation is to avoid confusion of
calling galaxies with exponential light profiles strictly as disk
galaxies, since dEs tend to have exponential profiles (see GG03).  The
distribution of these galaxies in the $L - \sigma$ plot (Figure 4,
right) indicates that there is a slight difference in the slopes for
each type.  Although a different $L - \sigma$ relation can be derived
for each galaxy type, the individual relations are still within
$3\sigma$ of each other (Table 2).  It is not surprising that a
significant number of single exponential component galaxies appear to
be present in this sample since this light profile best describes the
dwarf elliptical galaxies.  However, classifying these low-luminosity
galaxies is difficult without resolved photometry and we also note
discrepancies in classification depending on the literature source.
Nonetheless, the slopes of all three galaxy types, single exponential
component, bulge+exponential component and bulge-dominated galaxies,
are still in disagreement with the FJ relation, but consistent with $L
\sim \sigma^{2.01 \pm 0.36}$ as derived earlier for the low-luminosity
early--type galaxies.

Tonry (1981) attributed the change of slope in the $L-\sigma$ relation
to less luminous systems having significant rotation.  Subsequently,
Davies \textit{et al.}~(1983) investigated rotational properties of
about a dozen faint Es with $-20.5<M_B<-18$ mag ($-22.17<M_R<-19.67$
mag), and showed that the faint Es rotate more rapidly than most of
the bright Es.  More recent studies of the low-luminosity early--type
galaxies, i.e. Simien, \& Prugniel (2001), Geha, Guhathakurta, \& van
der Marel (2003), and van Zee, Skillman, \& Haynes (2004) indicate
that some of these objects have a rotational component while others
show little or no rotation.

We investigate whether rotational effects in the faint early--type
galaxies can cause the change of slope in the $L-\sigma$ relation
using the following approach.  We assume that there is a universal
relation between the luminosity and the kinetic energy per unit mass
(or KE) such as: $L \propto KE^2$, which is common to both faint
early--type galaxies and bright ellipticals.  Since the kinetic energy
per unit mass for a spheroid is (from Busarello, Longo \& Feoli 1992):
\begin{displaymath}
\mathrm{KE}=\frac{1}{2} \langle v^2 \rangle = \frac{1}{2} V_{rot}^2 +
\frac{3}{2} \sigma^2
\end{displaymath}
the assumed universal relation between luminosity and kinetic energy
per unit mass translates into a correlation between luminosity,
$\sigma$, and the anisotropy parameter ($V_{rot}/ \sigma$):
\begin{displaymath}
L=a \sigma^4 (\frac{V_{rot}^2}{\sigma^2} + 3)^2 
\end{displaymath}
For bright ellipticals, the existence of the Faber-Jackson relation
($L=b \sigma^4$) implies that $V_{rot}=0$, as it is indeed the case since
bright ellipticals are anisotropy-supported stellar systems and show
no or little rotation. For faint early--type galaxies, however, the
existence of a relation such as $L=c \sigma^2$, implies that there is a
systematic increase in the anisotropy parameter (i.e., in the amount
of rotation) as the velocity dispersion (or luminosity) decreases:
\begin{displaymath}
\frac{V_{rot}^2}{\sigma^2} = \sqrt{\frac{c}{b}} \frac{1}{\sigma} - 3
\end{displaymath}
This expression yields the amount of rotation a faint early--type galaxy
would need to have in order to follow the same relation between
luminosity and $\langle v^2\rangle$ followed by bright Es. Under these
assumptions, the observed change of slope in the $L-\sigma$ diagram would
simply be the result of not including the rotational component in the
kinematic energy of faint early--type systems.  

Using the expression above we can calculate the expected $V_{rot}/
\sigma$ for galaxies at different luminosities and velocity
dispersions, and compare them to the observed values.  We show this
comparison through a set of graphs (see Figure 6).  The left panel
shows $V_{rot}/ \sigma$ vs. $M_R$, and the right panel $V_{rot}/
\sigma$ vs. $\sigma$.  The solid line in both panels represents
the predicted value for $V_{rot}/ \sigma$, while the points are the
most recent data from the literature (Davies \textit{et al.}~1983;
Simien \& Prugniel 2002; Pedraz \textit{et al.}~2002; Geha,
Guhathakurta \& van der Marel 2003; van Zee, Skillman, \& Haynes
2004).

According to the predicted relation between the $V_{rot}/ \sigma$ and
$M_R$, $V_{rot}/ \sigma$ would have to increase steadily toward the
faint end (left panel of Figure 6).  For example, a faint early--type
galaxy with $M_R=-18$ mag and $\sigma=39$ km s$^{-1}$ would have $V_{rot}/
\sigma = 3.2$, or $V_{rot}=123$ km s$^{-1}$.  This value seems
unreasonably large when compared to observed $V_{rot}/ \sigma$ values
from Geha, Guhathakurta \& van der Marel (2003), which range from as
low as 0.01 to $\sim0.5$. The discrepancy is worse for the faintest
galaxies. We conclude that the predicted $V_{rot}/ \sigma$ is
inconsistent with observations of early--type galaxies fainter than
$M_R=-20.5$ mag.  Therefore, it is implausible that rotation is solely
responsible for the difference in the $L - \sigma$ slope between faint
early--type and bright elliptical galaxies.

An independent confirmation of $L \propto \sigma^2$ has recently been
provided by De Rijcke \textit{et al.}~(2004).  They investigate how
well different galaxy formation scenarios reproduce this slope
difference between the bright ellipticals and bulges of spirals and
dEs. The semi-analytical models which include quiescent star
formation, post-merger star-bursts and gas-loss triggered by supernova
winds seem to describe this effect well.

Next, we address the scatter around the $L-\sigma$ relation for the
faint early--type galaxies.  The rms scatter of the bisector line is
0.52 mag, where 0.22 mag can be attributed to the observational
errors, assuming $\delta M_R=0.1$ mag.  The scatter in the $L-\sigma$
relation of faint early--type galaxies is, therefore, intrinsic.  A
possible reason for this scatter could be that the faint early--type
galaxies in question are not yet relaxed.  This could indicate that in
their recent history these galaxies have encountered interactions with
other galaxies in the cluster.  It is also possible that the age of
the galaxy could have an effect on the scatter, as investigated by
Forbes, \& Ponman (1999).  We do not have direct age measurements for
these objects yet.  However, in the following section we plot the
colour$-\sigma$ relation for our galaxies.  Under assumptions
described in the following section we are able to investigate the
effects age and metallicity have on this relation.

In conclusion, faint early--type galaxies follow a well-defined $L
\propto \sigma^{2.01 \pm 0.36}$ relation.  This relation is distinct
from the traditional Faber-Jackson relation defined for bright E
galaxies and might indicate that bright Es should no longer be viewed
as canonical early--type galaxies.  We also conclude that rotation in
the faint early--type galaxies is not responsible for the change in the
slope relative to that derived for bright Es.

\subsection{Colour-$\sigma$ Relation}

The colour-magnitude relation (CMR) is a well established relation for
early--type galaxies.  It is characterized by the more luminous
galaxies displaying redder colours.  This relation was investigated by
many authors in the past who determined that the slope of the relation
arises because the more massive galaxies are redder and more metal
rich than the less massive galaxies (e.g., Terlevich, Caldwell, \&
Bower 2001, and references therein). Caldwell (1983) and Prugniel
\textit{et al.}~(1993) found that faint early--type galaxies roughly
follow the CMR for bright Es. A similar, distance-independent,
relation is the correlation between colour and $\sigma$, $C-\sigma$.
    
In Figure 5, we show the $C-\sigma$ relation for faint early--type
galaxies and bright ellipticals.  All of the galaxies seem to follow
the same relation, although we note the lack of bright Es in the plot
containing only Coma galaxies (left panel).  We confirmed the
uniformity of the $C-\sigma$ relation by checking our result with the
U--V colours of Terlevich, Caldwell, \& Bower (2001).  The right panel
includes galaxies from different clusters (Faber \textit{et al.}~1989)
and indicates that both faint and bright early--type galaxies follow
the same $C-\sigma$ relation.

Since luminosity and $\sigma$ are related, the $C-\sigma$ relation is
equivalent to the CMR, which in turn suggests a more fundamental
relation between galaxy metallicity and mass.  Although colours depend
both on metallicity and age changes in the stellar populations, the
evidence so far supports that metallicity changes are responsible for
the slope of the CMR, while age differences contribute to the scatter
observed around that relation.  In the following analysis we assume
that the same applies to the $C-\sigma$ relation.  Note however, that
it is likely that both age and metallicity affect the slope and the
scatter.  Bernardi \textit{et al.}~(2005) show that galaxies with
large velocity dispersion tend to be older.  They also show that at a
specific $\sigma$, galaxies have a wide range in both age and
metallicity in a way that the older galaxies have smaller
metallicities and the younger galaxies larger metallicities.  Assuming
that the intrinsic scatter in the $C-\sigma$ relation is predominantly
due to age, it is possible to constrain the age variations at a
given formation epoch for faint early--type galaxies.

Bower, Lucey and Ellis (1992) (hereafter BLE92) showed that it is
possible to determine minimum ages for galaxies with different
formation scenarios by implementing evolutionary stellar population
synthesis models and the intrinsic scatter in the CMR of these
galaxies.  We closely follow their method but use the intrinsic
scatter derived from the $C-\sigma$ relation for our galaxies instead.
Since the uncertainty in the observed parameters is 0.024 mag, the
intrinsic scatter is 0.067.

We used the evolutionary stellar population synthesis models of
Bruzual, \& Charlot (2003) to simulate a galaxy with an exponentially
declining star-burst, $\tau=1$ Gyr, and a Chabrier IMF.  In Figure 7,
top panel, we show how the $B-R$ colour of a galaxy with metalicitiy $Z=1$
or $0.4$ $Z_{\odot}$ evolves with time.  Hence, we find how the rate
of change of colour varies with the age of the galaxy (middle panel).
The colour change can be related to the intrinsic scatter in the colour
and the range of epochs for major star-formation events by the
relation:
\begin{equation}
\label{error_estimates}
{\sigma_{(B-R)}=\frac{d(B-R)}{dt} \times {\sigma_{SFE}}}
\end{equation}
where $\sigma_{(B-R)}$ is the scatter in $B-R$ colour and
$\sigma_{SFE}$ is the range in the star formation epoch (BLE92).  This
is in accordance with the assumption that the scatter in $C-\sigma$ is
only due to age variation, or in this case, to the scatter in the star
formation epoch.

After finding the d(B-R)/dt at different ages of the galaxy, and using
the intrinsic scatter of 0.067 in the $C-\sigma$ relation, we derive
values for the maximum range in the star formation epoch.  In Figure 7
(bottom panel) we show the maximum range in the star formation epochs
as a function of the formation time, constrained by the intrinsic
scatter in $C-\sigma$ for our sample.  If the average age of our
galaxy sample is 10 Gyr old, for example, the maximum range in its
star formation epoch will be $\sim 3$ Gyr.  However, in order to
determine the upper limit on variations in the ages of galaxies we
must take into account that the scatter in the SF epoch will also
depend on how the galaxies were formed.  Different formation scenarios
can be parametrised by a parameter $\beta$, which describes the degree
of coordination of galaxies during their formation.  BLE92 define
$\beta$ as `the ratio of the spread in formation time to the
characteristic time-scale at formation,' where the galaxy formation
times range as $\beta(t_H-t_F)$ ($t_H$ is the age of the universe, and
$t_F$ is the time of formation of the galaxy).  For example, $\beta=1$
for uncoordinated galaxy formation, and $\beta=0.1$ for strong
coordination.  Assuming solar metallicity, the minimum average age for
the faint early type galaxies in our sample with $\beta=0.1$ is
$\sim6$ Gyr (Figure 7, middle panel) with a scatter in the star
formation epoch of $\pm1$ Gyr (where $t_H=13.7$ Gyr, $\Omega_M=0.3$
and $\Omega_{\Lambda}=0.7$).  As a reference, a galaxy that is 6 Gyr
old must have formed at redshift $z \sim 0.7$.  Note that we can only
put a lower limit on the ages of galaxies and an upper limit on the
scatter in their SF epoch, provided that we know the level of
coordination of galaxies during their formation.

Although we initially assumed that the scatter of the $C-\sigma$
relation is due to an age spread around the formation epoch in a
single burst, secondary bursts of star formation will also contribute
to the scatter.  In fact, observations of galaxies in clusters at
redshifts $z \sim 0.5$ point to a possibility of secondary star-bursts
(Butcher \& Oemler 1978; Butcher \& Oemler 1984) in what may become
today's population of faint early--type galaxies in clusters.  By
modeling the secondary star-bursts, we can place upper limits on the
star-burst strengths.  

Assuming that all galaxies formed at $t>10$ Gyrs and had a secondary
starburst $\sim 5$ Gyrs ago, we again use the models of Bruzual, \&
Charlot (2003) with two exponentially declining bursts to make this
constraint.  For this purpose we also follow the discussion of BLE92.
Using the scatter in the $B-R$ colour for various uniformly
distributed burst strengths, solar metallicity and a Chabrier IMF, we
find the typical rms burst strength $r_{typ}= \Delta (B-R)/0.17$ (the
ratio between the stellar mass of the secondary burst to that of the
initial burst).  Our observed scatter of $(B-R)_{rms} = 0.07$ places
an upper limit on the secondary burst of 40 percent by stellar mass of
the first burst.  Unfortunately, this constraint is not very robust
since it strongly depends on the assumptions of the age of the
starburst and the modeling components.

In conclusion, we have shown that there is a well defined relation
between colour and $\sigma$ for faint early--type systems. Assuming that
metallicity changes are responsible for the slope of this correlation
while age variations are the main contributor to the scatter, it is
possible to constrain the age range of major star formation events for
a given formation epoch. However, it is difficult to decouple the
effects of age and metallicity using colours. In future papers we will
study the detailed stellar population properties of faint early--type
galaxies, both age and metallicity, using line strength indices and
stellar population synthesis models.  Furthermore, we will test if the
galaxies in the centre of the cluster are more metal-rich than those
in the outskirts, since this is predicted by the galaxy harassment
model (Moore, Lake, \& Katz 1998).

\section{Summary}

We present velocity dispersion measurements for 69 faint early--type
galaxies in the centre of the Coma cluster with $-22.17 \lea M_R \lea
-17.5$ mag.  We derive the $L-\sigma$ relation for faint early--type
galaxies as $L \propto \sigma^{2.01\pm0.36}$, which differs from the
Faber-Jackson relation, $L \propto \sigma^4$, defined for bright
ellipticals.  Rotation in these objects is investigated as a possible
cause for the difference in the slope.  Although rotation may
contribute to the scatter in this relation, it is not the main cause
for the different slope derived for these galaxies.

We also investigate whether the slope change is due to the presence of
different classes of early--type galaxies in our sample.  Although our
sample includes bulge-dominated, bulge+single exponential component
and a few single exponential component galaxies, all three types
essentially follow the same relation.

We find that faint early--type galaxies follow a well-defined
$C-\sigma$ relation.  By assuming that this relation is mostly driven
by an increased metallicity with increasing galaxy mass, while the
scatter reflects age differences, we investigated how we can constrain
either the ages, the range of star formation epoch, or the strength of
secondary bursts for the faint early--type galaxies for various galaxy
formation scenarios.

\section*{Acknowledgments}

We thank our referee, Michael Hudson, for his suggestions which have
significantly improved a previous version of this paper.  We also
thank Alister Graham for many insightful discussions, Javier Gorgas,
Nicol\'as Cardiel and Patricia S\'anchez-Bl\'azquez for help and
instructions on velocity dispersion measurements; Marla Geha for an
independent check of velocity dispersions measurements; Neil Trentham
for providing the B-R colours for our galaxies; Eric McKenzie for
discussions and help with stellar synthesis models; Nicolas Gruel for
help and use of his catalogue cross-correlation program; and Nelson
Caldwell for providing the U-V coluor catalog.  R. G. acknowledges
funding from the archive HST proposal HF01092.01-97A.  R.G. also
thanks the Yale TAC for generous allocation of time on the WIYN
telescope.


\clearpage

\begin{table*}
\centering
\begin{minipage}{160mm}
\caption{Table of $\sigma$ measurements}
\begin{tabular}{@{}llllllrrrr@{}}
\hline 
\hline
GMP & Galaxy &RA &Dec. & b$_j$ & V$_r$ & $\delta V_r$ & $\sigma$ & $\delta \sigma$ & SNR \\ 
No.& Name& h:m:s& $^\circ$ $\arcmin$ $\arcsec$& mag& km s$^{-1}$& km s$^{-1}$& km s$^{-1}$& km s$^{-1}$  \\ 
(1)&  (2)&   (3)&   (4) & (5) & (6)& (7) & (8)& (9)& (10) \\ 
\hline
2478 &         &13:00:45.46 &27:50:06.00 &18.09 &8670.5 &2.0  &50.0  &5.1  & 28  \\
2489 &         &13:00:44.69 &28:06:00.70 &16.69 &6539.1 &1.3  &93.8  &2.3  & 54  \\
2510 &         &13:00:42.92 &27:57:45.49 &16.13 &8336.9 &2.0  &126.7 &2.5  & 48  \\
2516 &IC 4042  &13:00:42.85 &27:58:14.90 &15.34 &6321.3 &3.6  &176.2 &4.6  & 46  \\
2519 &         &13:00:42.65 &28:06:56.90 &18.68 &6107.0 &26.5 &71.2  &76.5 & 13  \\
2529 &         &13:00:41.28 &28:02:40.74 &18.63 &8519.6 &2.1  &42.3  &5.5  & 28  \\
2535 &IC 4041  &13:00:40.94 &27:59:46.19 &15.93 &7026.9 &2.0  &123.8 &2.4  & 47  \\
2541 &NGC 4906 &13:00:39.83 &27:55:24.40 &15.44 &7436.6 &3.9  &176.6 &4.0  & 45  \\
2585 &         &13:00:35.50 &27:56:32.15 &18.44 &6898.7 &2.6  &29.6  &6.3  & 23  \\
2603 &         &13:00:33.45 &27:49:25.60 &17.36 &8099.9 &1.6  &53.7  &4.7  & 39  \\
2654 &         &13:00:28.06 &27:57:19.75 &16.38 &6940.7 &2.5  &142.7 &3.5  & 51  \\
2676 &         &13:00:26.24 &28:00:30.17 &19.03 &5527.8 &6.7  &48.9  &18.8 & 10  \\
2692 &         &13:00:24.90 &27:55:34.14 &18.20 &7866.3 &2.7  &41.4  &5.6  & 26  \\
2778 &         &13:00:18.88 &27:56:11.75 &16.69 &5260.6 &1.5  &56.8  &3.9  & 34  \\
2784 &         &13:00:18.62 &28:05:48.01 &18.36 &7739.8 &6.3  &63.8  &7.5  & 15  \\
2799 &         &13:00:17.72 &27:59:13.29 &18.70 &5968.0 &4.4  &44.6  &9.4  & 17  \\
2805 &         &13:00:17.11 &28:03:48.29 &16.57 &6103.7 &2.1  &128.9 &2.5  & 48  \\
2839 &IC 4021  &13:00:14.83 &28:02:26.93 &16.01 &5720.8 &2.6  &160.0 &2.8  & 53  \\
2852 &         &13:00:13.72 &27:52:00.28 &17.80 &7327.5 &1.8  &41.5  &3.0  & 27  \\
2861 &         &13:00:12.98 &28:04:30.05 &16.26 &7455.4 &1.8  &123.2 &2.3  & 53  \\
2879 &         &13:00:11.23 &28:03:53.16 &18.05 &7270.6 &2.5  &43.8  &5.2  & 23  \\
2912 &NGC 4895A&13:00:09.20 &28:10:11.73 &16.07 &6713.3 &2.7  &149.9 &2.8  & 54  \\
2917 &         &13:00:08.50 &27:57:15.23 &19.13 &6212.6 &7.0  &51.8  &13.1 & 12  \\
2922 &IC 4012  &13:00:08.10 &28:04:41.15 &15.93 &7185.7 &3.6  &185.1 &4.2  & 45  \\
2929 &         &13:00:07.56 &27:57:27.22 &18.66 &6178.2 &6.7  &22.4  &15.5 & 13  \\
2940 &IC 4011  &13:00:06.48 &28:00:13.19 &16.08 &7182.8 &2.9  &126.3 &4.2  & 45  \\
2960 &         &13:00:05.47 &28:01:26.46 &16.78 &5847.6 &1.6  &68.7  &2.4  & 45  \\
2985 &         &13:00:03.83 &27:57:51.31 &17.87 &5310.7 &4.8  &43.1  &10.2 & 15  \\
3017 &         &13:00:01.05 &27:56:41.80 &17.91 &6790.9 &4.7  &58.5  &12.2 & 22  \\
3018 &         &13:00:01.08 &27:59:27.80 &19.31 &7476.1 &21.0 &75.7  &71.2 & 11  \\
3034 &         &12:59:59.56 &27:56:24.42 &18.06 &6106.0 &26.0 &169.4 &84.4 &  9  \\
3058 &         &12:59:57.72 &28:03:52.35 &17.71 &5791.0 &2.8  &38.0  &7.8  & 23  \\
3068 &         &12:59:56.75 &27:55:46.40 &16.47 &7646.6 &2.2  &106.3 &2.9  & 44  \\
3073 &NGC 4883 &12:59:56.10 &28:02:03.43 &15.43 &8054.0 &3.8  &173.3 &5.2  & 45  \\
3098 &         &12:59:54.03 &27:58:12.05 &18.63 &6740.7 &7.3  &59.2  &35.0 & 14  \\
3121 &         &12:59:51.53 &28:04:22.80 &17.34 &7405.6 &3.0  &49.8  &6.2  & 24  \\
3126 &         &12:59:51.08 &27:49:56.68 &17.55 &7826.1 &2.1  &53.7  &3.8  & 26  \\
3131 &         &12:59:50.26 &27:54:43.52 &18.68 &7195.5 &5.0  &49.9  &8.5  & 13  \\
3133 &         &12:59:50.18 &27:55:27.65 &17.23 &9674.8 &2.0  &79.6  &3.0  & 35  \\
3166 &         &12:59:47.03 &27:59:29.02 &18.37 &8315.3 &5.1  &58.8  &8.0  & 16  \\
3170 &IC 3998  &12:59:46.88 &27:58:24.04 &15.70 &9302.1 &2.2  &142.6 &4.2  & 41  \\
3196 &         &12:59:44.77 &27:53:21.47 &18.35 &6747.3 &3.9  &55.0  &10.3 & 19  \\
3201 &NGC 4876 &12:59:44.47 &27:54:43.02 &15.51 &6657.0 &2.8  &169.7 &3.6  & 45  \\
3205 &         &12:59:44.30 &27:52:01.98 &17.61 &6203.7 &3.3  &56.8  &3.5  & 25  \\
3206 &         &12:59:44.29 &27:57:28.36 &16.36 &6856.5 &2.1  &79.6  &3.3  & 33  \\
3209 &         &12:59:44.25 &28:00:45.14 &19.37 &7097.4 &7.6  &66.5  &14.2 & 15  \\
3213 &         &12:59:43.80 &27:59:39.15 &16.14 &6636.4 &3.4  &138.7 &5.1  & 48  \\
3222 &         &12:59:42.38 &27:55:27.37 &16.47 &6866.2 &3.8  &163.6 &4.1  & 41  \\
3254 &         &12:59:40.36 &27:58:03.95 &16.57 &7520.9 &3.0  &100.5 &4.9  & 28  \\
3269 &         &12:59:39.73 &27:57:12.57 &16.12 &7947.8 &2.2  &98.5  &2.7  & 38  \\
3292 &         &12:59:38.08 &28:00:01.79 &17.70 &4954.9 &2.6  &63.8  &6.0  & 22  \\
3296 &         &12:59:38.00 &27:54:24.64 &15.88 &7922.5 &3.5  &179.8 &4.0  & 49  \\
3312 &         &12:59:37.09 &28:01:05.15 &18.68 &7156.1 &3.3  &46.5  &8.4  & 25  \\
3313 &         &12:59:37.06 &27:49:31.05 &17.53 &6191.3 &1.9  &85.2  &2.9  & 49  \\
3336 &         &12:59:35.58 &27:54:19.86 &18.47 &6954.3 &3.3  &52.3  &5.1  & 19  \\
3339 &         &12:59:35.37 &27:51:47.40 &17.54 &6261.5 &2.1  &58.5  &4.4  & 33  \\
3340 &         &12:59:35.25 &27:56:03.28 &18.54 &4496.2 &18.2 &63.0  &48.8 & 11  \\
3352 &NGC 4872 &12:59:34.22 &27:56:47.18 &14.79 &7120.1 &4.3  &211.1 &6.4  & 42  \\
3367 &NGC 4873 &12:59:32.86 &27:58:59.55 &15.15 &5804.8 &3.6  &178.9 &3.5  & 45  \\
\hline
\end{tabular}
\end{minipage}
\end{table*}

\setcounter{table}{0}
\begin{table*}
\centering
\begin{minipage}{160mm}
\caption{{\sl ctd.} Table of $\sigma$ measurements}
\begin{tabular}{@{}llllllrrrr@{}}
\hline 
\hline
GMP & Galaxy &RA &Dec. & b$_j$ & V$_r$ & $\delta V_r$ & $\sigma$ & $\delta \sigma$ & SNR \\ 
No.& Name& h:m:s& $^\circ$ $\arcmin$ $\arcsec$& mag& km s$^{-1}$& km s$^{-1}$& km s$^{-1}$& km s$^{-1}$  \\ 
(1)&  (2)&   (3)&   (4) & (5) & (6)& (7) & (8)& (9)& (10) \\ 
\hline
3376 &         &12:59:32.19 &27:55:14.35 &18.24 &7041.4 &11.5 &46.4  &27.3 & 11  \\
3400 &IC 3973  &12:59:30.90 &27:53:01.74 &15.32 &4696.2 &4.0  &216.5 &4.9  & 51  \\
3438 &         &12:59:28.60 &28:01:07.71 &19.01 &5942.2 &4.1  &32.7  &10.9 & 15  \\
3471 &         &12:59:26.67 &27:59:52.73 &16.45 &6629.2 &1.8  &86.5  &2.5  & 49  \\
3486 &         &12:59:25.41 &27:56:02.48 &17.73 &7522.7 &2.1  &57.2  &2.7  & 32  \\
3510 &NGC 4869 &12:59:23.42 &27:54:39.71 &14.97 &6791.6 &3.9  &195.5 &4.5  & 46  \\
3534 &         &12:59:21.51 &27:58:23.01 &17.20 &5994.6 &1.2  &53.5  &2.9  & 43  \\
3554 &         &12:59:20.28 &28:04:25.70 &17.20 &7091.3 &1.5  &62.6  &3.2  & 43  \\
3561 &NGC 4865 &12:59:19.87 &28:05:01.70 &14.54 &4499.2 &3.7  &214.5 &3.5  & 58  \\
3565 &         &12:59:19.79 &27:58:22.80 &18.44 &7172.6 &3.6  &46.4  &12.5 & 15  \\
3625 &         &12:59:15.91 &27:53:07.67 &19.63 &6521.8 &15.4 &108.3 &24.3 &  8  \\
3629 &         &12:59:15.66 &27:53:55.19 &19.03 &5219.2 &13.8 &44.6  &28.2 &  5  \\
3645 &         &12:59:14.71 &27:53:42.42 &18.64 &6367.3 &2.9  &69.4  &7.7  & 25  \\
3656 &         &12:59:14.02 &28:04:32.70 &15.53 &7729.2 &2.5  &142.1 &4.6  & 53  \\
3681 &         &12:59:11.64 &28:00:31.46 &18.01 &6818.4 &3.0  &64.2  &5.4  & 21  \\
3706 &IC 3960A &12:59:09.73 &27:52:00.86 &17.61 &6851.9 &1.8  &95.7  &2.5  & 49  \\
3707 &         &12:59:09.53 &28:02:25.56 &17.76 &7150.9 &2.1  &76.8  &3.6  & 38  \\
3733 &IC 3960  &12:59:08.04 &27:51:16.13 &15.85 &6520.2 &3.2  &175.9 &6.0  & 45  \\
3739 &IC 3957  &12:59:07.58 &27:46:02.22 &15.88 &6316.9 &2.4  &172.5 &4.1  & 46  \\
3761 &IC 3955  &12:59:06.11 &27:59:46.65 &15.57 &7602.9 &2.9  &153.8 &3.0  & 49  \\
3780 &         &12:59:04.87 &28:02:59.94 &17.89 &7957.3 &2.0  &57.9  &3.4  & 30  \\
3782 &         &12:59:04.73 &27:54:37.90 &16.55 &6363.0 &2.0  &110.7 &2.5  & 56  \\
3792 &NGC 4860 &12:59:03.99 &28:07:23.61 &14.69 &7844.4 &6.1  &256.9 &7.2  & 41  \\
3794 &         &12:59:04.24 &27:57:31.19 &17.37 &6952.6 &2.5  &130.0 &3.1  & 47  \\
3851 &         &12:59:00.15 &27:58:01.08 &16.98 &8233.8 &1.8  &83.2  &2.7  & 41  \\
3855 &         &12:58:59.56 &27:56:02.57 &18.05 &5702.7 &4.7  &49.7  &5.9  & 22  \\
3856 &         &12:58:59.59 &27:59:34.47 &19.58 &6186.7 &32.0 &122.8 &37.4 &  8  \\
3914 &         &12:58:55.34 &27:57:51.26 &16.57 &5650.6 &3.7  &161.1 &3.6  & 46  \\
\hline
\end{tabular}
\\Table 1:  Notes.--Col.(1).--Galaxy number according to Goodwin, Metcalfe, \&
Peach (1983), (GMP from here on).\\
Col.(2).--NGC or IC name of galaxy.\\
Col.(3).--RA\\
Col.(4).--Dec\\
Col.(5).--$b_j$ magnitude from the GMP catalog\\
Col.(6).--Radial velocity, $V_r$. \\
Col.(7).--Uncertainty in the radial velocity measurement, $\delta V_r$.\\
Col.(8).--Velocity dispersion measurement, $\sigma$\\
Col.(9).--Uncertainty in the velocity dispersion, $\delta \sigma$.\\
Col.(10).--Signal-to-noise ratio of galaxies.\\
\end{minipage}
\end{table*}
\clearpage

\begin{table*}
\begin{minipage}{140mm}
\caption{$L-\sigma$ Least-Squares Fits}
\begin{tabular}{@{}lccccc@{}}
\hline
\hline
Galaxy Type & Regression & Intercept & Slope & rms &  n\\
(1)&  (2)&   (3)&   (4) & (5) & (6) \\ 
\hline
all galaxies & $M_R\mid \log \sigma$ & -10.949$\pm$0.375 & -4.551$\pm$0.180 & 0.510 mag& 1.82$\pm$0.07\\  
all galaxies & $\log \sigma \mid M_R$ &  -8.846$\pm$0.444 & -5.585$\pm$0.210 & 0.565 mag& 2.23$\pm$0.08\\  
all galaxies & Bisector            & -10.001$\pm$0.376 & -5.017$\pm$0.179 & 0.521 mag& 2.01$\pm$0.07\\  
\hline
Bulge--dominated & Bisector &  -9.194$\pm$0.565 & -5.370$\pm$0.269 & 0.653 mag& 2.15$\pm$0.11\\
Bulge+Exp. Comp.     & Bisector & -11.374$\pm$0.608 & -4.370$\pm$0.306 & 0.464 mag& 1.75$\pm$0.12\\
Exponential Comp. & Bisector & -12.021$\pm$0.687 & -4.101$\pm$0.319 & 0.628 mag& 1.64$\pm$0.13\\
\hline
\end{tabular}
\\Table 2: Notes.--Col.(1).--Galaxy type in regression \\
Col.(2).--Regression order: $M_R\mid \log \sigma$, means minimizing in $M_R$ on
$\log \sigma$. In case of Single Exponential Component, Bulge+Single
Exponential Component and Bulge--dominated galaxies we only show the
bisector value (Figure 4, {\it right} pannel).\\ 
Col.(3).--Intercept and uncertainty in linear regression \\ 
Col.(4).--Slope and uncertainty in linear regression \\
Col.(5).--rms of points around the linear fit. \\ 
Col.(6).--The power $n$ of $L \propto \sigma^n$. \\
\\
\\
\\
\\
\\
\end{minipage}
\end{table*}

\begin{table*}
\begin{minipage}{140mm}
\caption{$C-\sigma$ Least Squares Fits}
\begin{tabular}{@{}ccccc@{}}
\hline
\hline
Panel & Regression & Intercept & Slope & rms\\
\hline
Left & B--R $\mid \log \sigma$   &   0.904$\pm$0.045 &  0.323$\pm$0.023 & 0.066 mag\\
     & $\log \sigma \mid$ B--R   &   0.435$\pm$0.162 &  0.555$\pm$0.081 & 0.066 mag\\ 
     & Bisector            &   0.680$\pm$0.089 &  0.434$\pm$0.045 & 0.071 mag\\
\hline
Right & B--R $\mid \log \sigma$   &   1.007$\pm$0.026 &  0.279$\pm$0.011 & 0.049 mag\\
      & $\log \sigma \mid$ B--R   &   0.595$\pm$0.080 &  0.463$\pm$0.036 & 0.049 mag\\ 
      & Bisector            &   0.807$\pm$0.041 &  0.369$\pm$0.018 & 0.052 mag\\
\hline
\end{tabular}
\end{minipage}
\end{table*}

\clearpage
\begin{figure}
\includegraphics{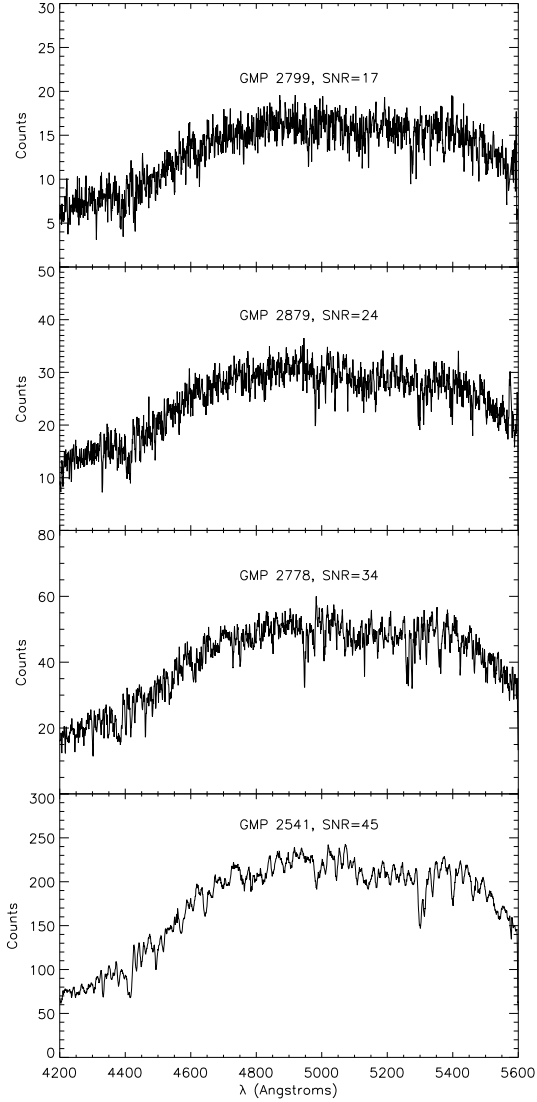}
\caption{Spectra for early type galaxies of different luminosities.
The GMP number (Goodwin, Metcalfe, \& Peach 1983) and the
signal-to-noise ratios (SNR) are given above each galaxy spectrum.
\label{fig1}}
\end{figure}

\begin{figure}
\includegraphics[width=9cm]{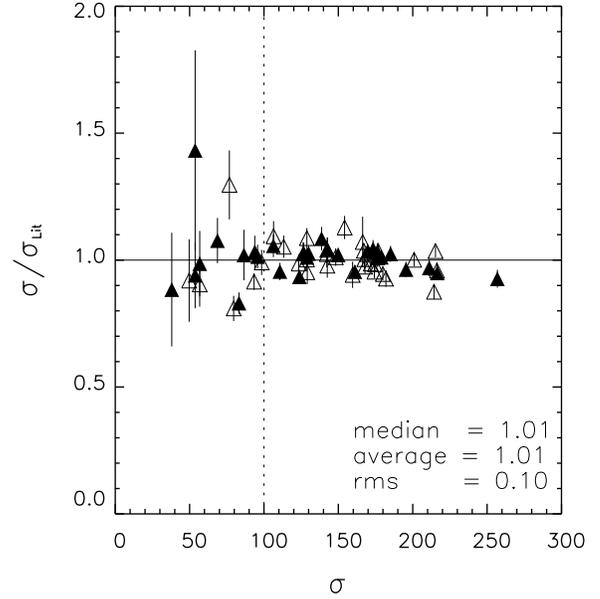}
\caption{External check of $\sigma$ measurements.  The plot shows our
$\sigma$ measurements compared to NFPS (open triangles) and MLKC02
(solid triangles).  The error bars in the $\sigma/ sigma_{Lit}$ are
also shown.
\label{fig2}}
\end{figure}

\begin{figure}
\includegraphics[height=8cm]{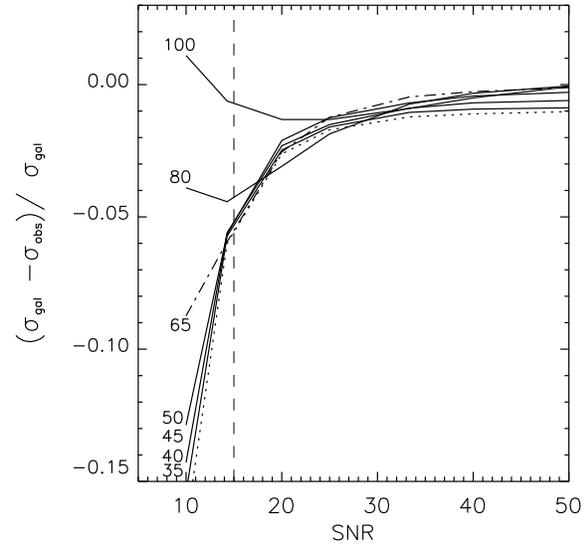}
\caption{Systematic effects on $\sigma$ measurements depending on the
SNR of the galaxy.  The y-axis shows the difference between the
spectra with a given velocity dispersion, $\sigma_{gal}$, and the
observed velocity dispersion, $\sigma_{obs}$, i.e. the same spectra
with different amount of noise.  The spectra were derived from
simulations.  The dotted line is for a spectrum of 35 km s$^{-1}$,
while dash--dot line is for 65 km s$^{-1}$.  The other lines are
marked accordingly.  We also included the dashed line at SNR of 15 as
we chose this value as the upper limit for reliable $\sigma$
measurements in this paper.
\label{fig3}}
\end{figure}
\clearpage

\begin{figure*}
\includegraphics[height=9cm]{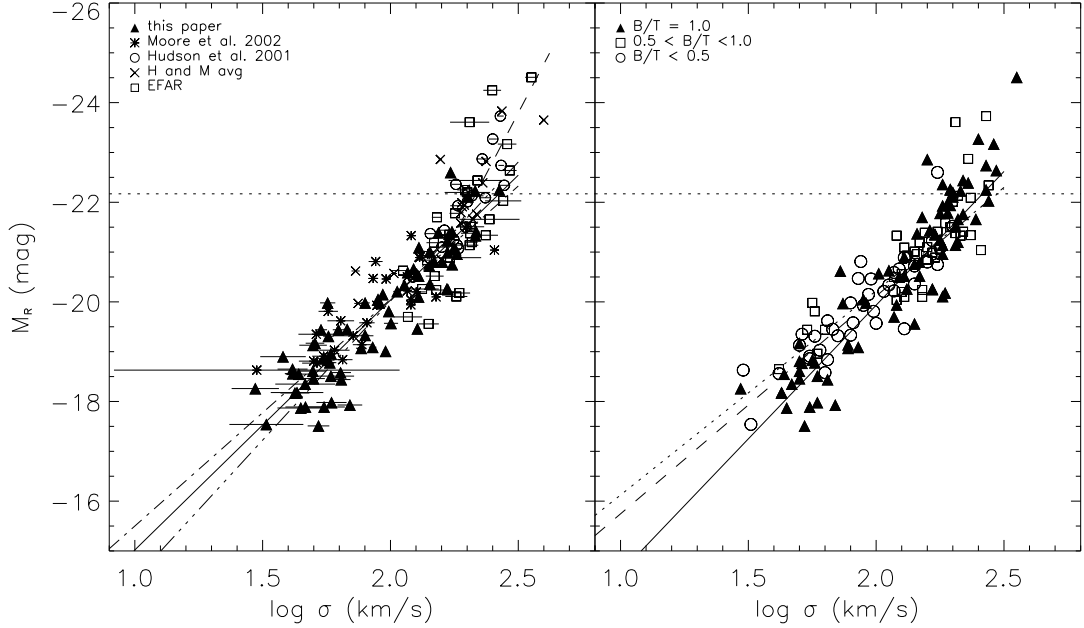}
\caption{$M_R-\sigma$ plots for: (Left) All the Coma galaxies in our
sample and those from the literature (see figure legend).  The {\it
dashed} line is the most recent FJ from literature, $L \propto
\sigma^{3.92}$ (Forbes \& Ponman, 1999); the {\it dash-dot} line is a
least squares fit when minimising the residuals in $M_R$; the {\it
dash-dot-dot} when minimizing in $\log \sigma$; and the {\it solid}
line is the bisector fit.  (Right) Including only galaxies for which
Guti\'errez \textit{et al.}~(2004) had the bulge-to-total ratio (B/T)
measurements.  The {\it solid} line is the least squares bisector fit
for bulge-dominated (solid triangles), the {\it dashed} line for
bulge+single exponential component (open squares), and the {\it
dotted} line for single exponential component (open circles) galaxies.
\label{fig4}}
\end{figure*}

\begin{figure*}
\includegraphics[height=9cm]{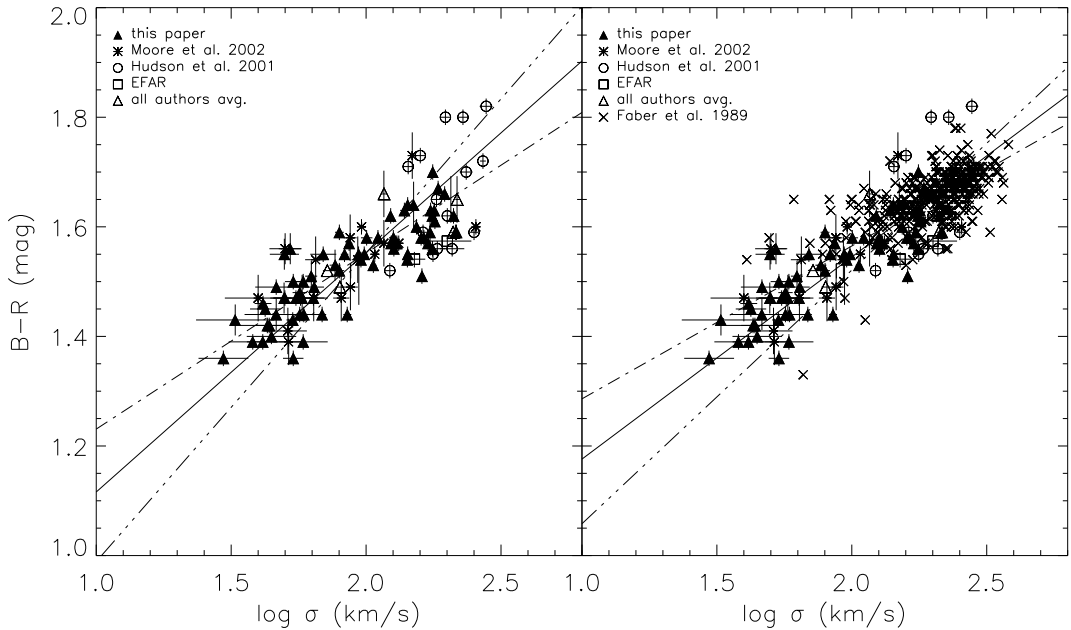}
\caption{$Colour-\sigma$ relation for early--type galaxies.  In both
panels the {\it dash-dotted} line represents a least squares linear
fit when minimizing the residuals in $B-R$, {\it dash-dot-dot} when
minimizing in $\log \sigma$, and the {\it solid} line is the bisector
line.  The {\it left} panel shows the $C-\sigma$ relation for the
galaxies in the Coma cluster, while the {\it right} includes galaxies
from all the clusters studied by Faber \textit{et al.} (1989).
\label{fig5}}
\end{figure*}
\clearpage

\begin{figure*}
\includegraphics{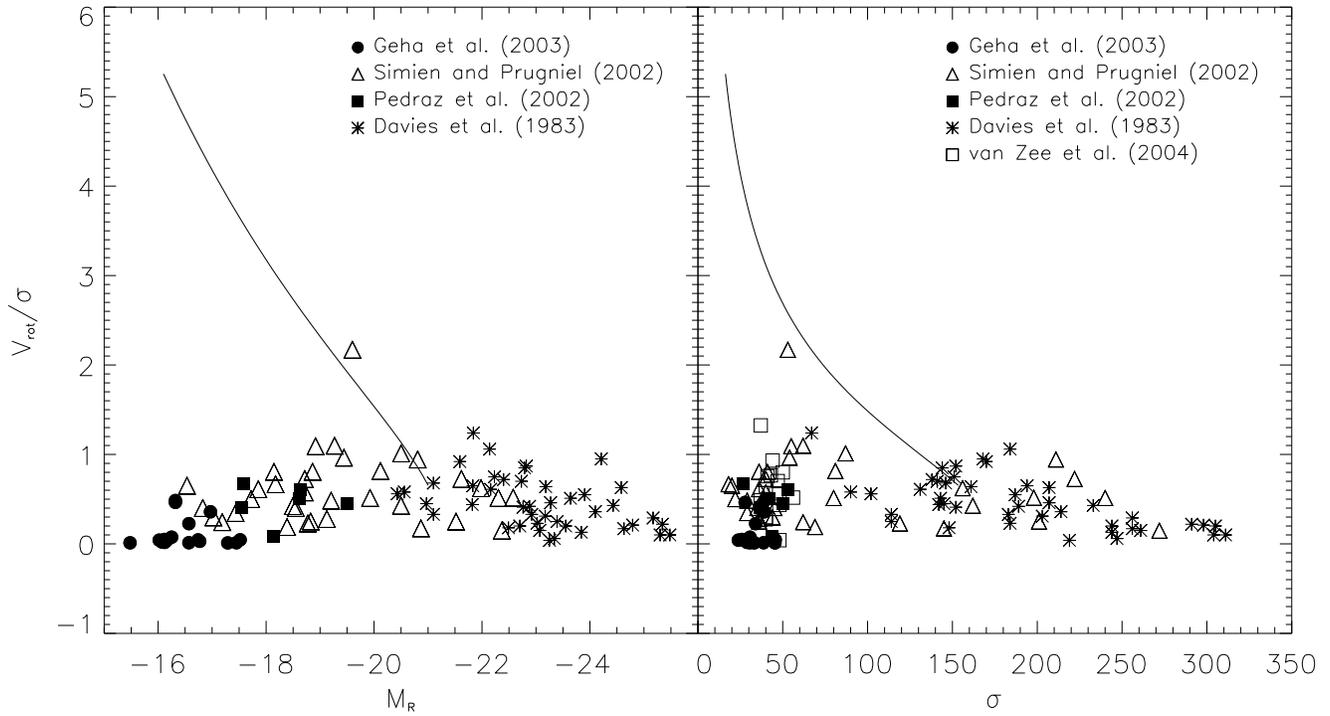}
\caption{Comparison between observed and predicted relation $V_{rot}/
\sigma$ with $M_R$ (Left) and $\sigma$ (Right).  In both panels the
solid line represents the predicted $V_{rot}/ \sigma$ a galaxy would
have to have in order for the faint early--type galaxies to follow the
$L\propto \sigma^4$ relation defined for bright Es.  The symbols are
as shown in the figure legend.
\label{fig6}}
\end{figure*}

\clearpage
\begin{figure}
\includegraphics{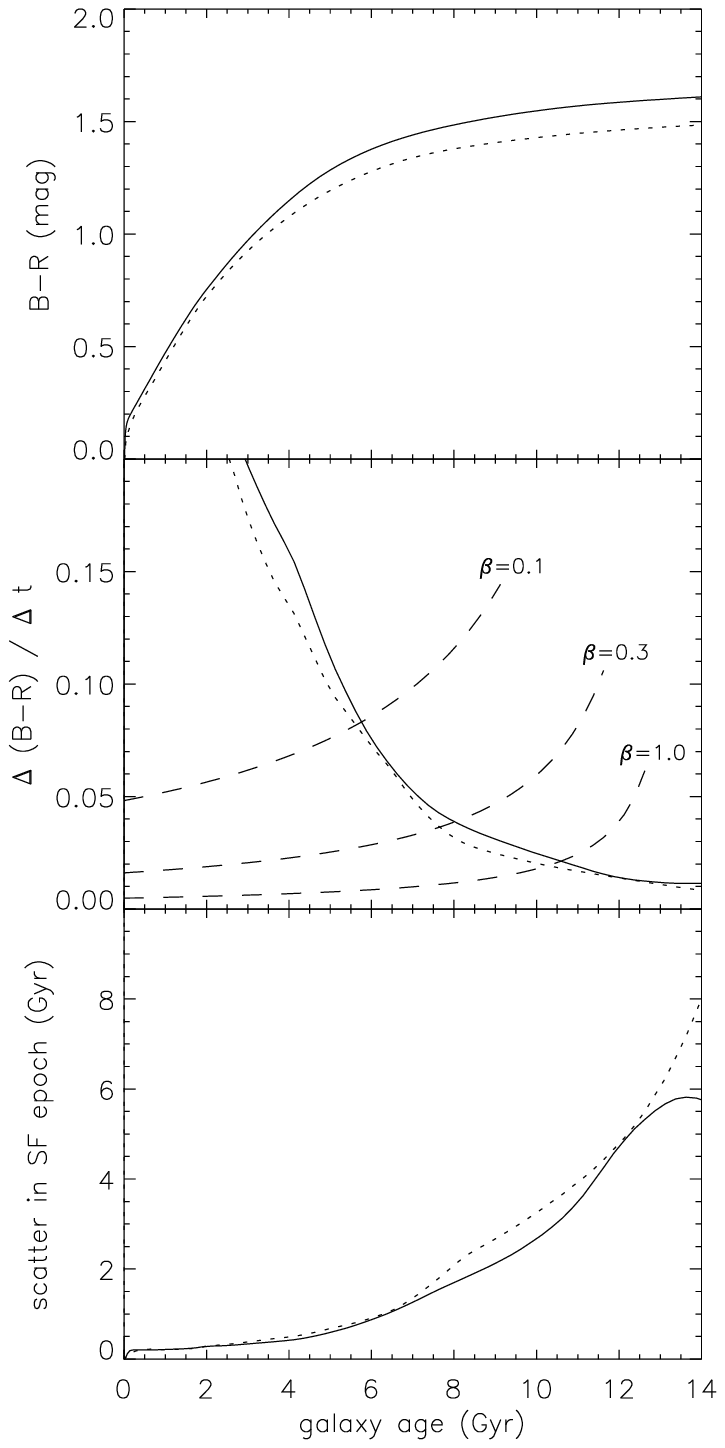}
\caption{(Top) $B-R$ colour vs. galaxy age, assuming exponential
burst.  The {\it solid} line represents a galaxy with $Z_{\odot}$,
while the {\it dotted} line is $0.4~Z_{\odot}$.  (Middle) The rate of
change of $B-R$ colour with time of formation.  The {\it dashed} lines
represent different $\beta$ parameters which depend on how
synchronized the galaxy formation is assumed to be.  $\beta$ of 1.0
corresponds to no coordination, while $\beta = 0.1$ is for strong
coordination.  (Bottom) Scatter in the star formation epoch vs. galaxy
age.  Assuming strong coordination in galaxy formation, the galaxy age
of 6 Gyrs (from middle pannel) would impliy that its scatter in SF
epoch is $\sim 1$ Gyr.
\label{fig7}}
\end{figure}

\bsp
\label{lastpage}


\begin{thebibliography}{}

\bibitem[]{} Bender, R., Burstein, D., \& Faber, S. M., 1992, ApJ, 399, 462
\bibitem[]{} Bender, R., \& Nieto, J. L., 1990, A\&A, 239, 97
\bibitem[]{} Bernardi, M., Renzini, A., da Costa, L.N., Wegner, G., Alonso, M.V., Pellegrini, P.S., Rité, C., Willmer, C.N.A., 1998, ApJL, 508, 143
\bibitem[]{} Bernardi, M., Sheth, R. K., Nichol, R. C., Schneider, D. P., \& Brinkmann, J., 2005, AJ, 129, 61 
\bibitem[]{} Binggeli, B., Sandage, A., \& Tarenghi, M., 1984, AJ, 89, 64
\bibitem[]{} Bower, R.G., Lucey, J.R., \& Ellis, R.S., 1992, MNRAS, 254, 601
\bibitem[]{} Brodie, J. P., \& Huchra, J. P., 1991, ApJ, 379, 157
\bibitem[]{} Bruzual, G., \& Charlot, S., 2003, MNRAS, 344, 1000
\bibitem[]{} Busarello, G., Longo, \& Feoli, A., 1992, A\&A, 262, 52
\bibitem[]{} Butcher, H., \& Oemler, A. Jr., 1978, ApJ, 219, 18
\bibitem[]{} Butcher, H., \& Oemler, A. Jr., 1984, ApJ, 285, 426
\bibitem[]{} Caldwell, N., 1983, AJ, 88, 804
\bibitem[]{} Caldwell, N., 1987, AJ, 94, 1116
\bibitem[]{} Caldwell, N., \& Bothun, G. D., 1987, AJ, 94, 1126
\bibitem[]{} Capaccioli, M., Caon, N., \& C'Onforio, M. 1992, MNRAS, 259, 323
\bibitem[]{} Cardiel, N. 1999, Dissertation, {\it Star formation in Central Cluster Galaxies}, Univ. Complutense, Madrid
\bibitem[]{} Colless, M., Dalton, G., Maddox, S., Sutherland, W., Norberg, P., Cole, S., Bland-Hawthorn, J., Bridges, T., Cannon, R., Collins, C., and 19 coauthors, 2001, MNRAS, 328, 1039
\bibitem[]{} Colless, M., \& Dunn, A.M., 1996, ApJ, 458, 435
\bibitem[]{} Davies, R.L., Efstathiou, G., Fall, M., Illingworth, G.,
\& Schechter, R.L., 1983, AJ, 266, 41
\bibitem[]{} de Carvalho, R. R., \& Djorgovski, S., 1992, ApJ, 389, 49
\bibitem[]{} De Rijcke, S., Dejonghe, H., Zeilinger, W. W., Hau,
G. K. T., 2001, ApJL, 559, 21
\bibitem[]{} De Rijcke, S., Michielsen, D., Dejonghe, H., Zeilinger, W. W., \& Hau, G. K. T., 2004, astroph 0412553, v2
\bibitem[]{} de Vaucouleurs, G., 1948, Ann. d'Astrophys., 11, 247
\bibitem[]{} Faber, S. M., \& Jackson, R. E., 1976, ApJ, 204, 668
\bibitem[]{} Faber, S. M., \& Lyn, D. N. C., 1983, ApJ, 266, 17
\bibitem[]{} Faber, S. M., Wegner, G., Burstein, D., Davies, R.,
Dressler, A., Lynden-Bell, D., \& Terlevich, R. J., 1989, ApJS, 69,
763
\bibitem[]{} Feigelson, E. D., \& Babu, G. J., 1992, AJ, 397, 55
\bibitem[]{} Ferguson, H. C., \& Binggeli, B., 1994, A\&A Rev., 6, 67
\bibitem[]{} Ferguson, H.C., \& Sandage, A., 1988, AJ, 96, 1520
\bibitem[]{} Forbes, D.A., \& Ponman, T.J., 1999, MNRAS, 309, 623
\bibitem[]{} Fukugita, M., Shimasaku, K, \& Ichikawa, T., 1995, PASP,
107, 945
\bibitem[]{} Geha, M., Guhathakurta, P., \& van der Marel, R., 2002,
AJ, 124, 3073
\bibitem[]{} Geha, M., Guhathakurta, P., \& van der Marel, R., 2003,
AJ, 126, 1794
\bibitem[]{} Gorgas, J., Pedraz, S., Guzm\'an, R., Cardiel, N., \&
Gonzalez, J. J., 1997, ApJ, 481, 19
\bibitem[]{} Gonz\'alez-Gonz\'alez, J. J., 1993, Dissertation, {\it
Line-Strength Gradients And Kinematic Profiles in Elliptical
Galaxies}, UC Santa Cruz
\bibitem[]{} Goodwin J., G., Metcalfe, N., \& Peach, J., V., (1983)
\bibitem[]{} Graham, A. W., \& Guzm\'an R., 2003, AJ, 125, 2936 
\bibitem[]{} Graham, A. W., 2004, ApJ, 613L
\bibitem[]{} Guti\'errez, C. M., Trujillo, I., Aguerri, J. A., Graham,
A. W.,\& Caon, N., \ 2004, ApJ, 602, 644
\bibitem[]{} Guzm\'an, R., Graham, A. W., Matkovi\'c A., Vass,
I. Gorgas,J., \& Cardiel, N., 2003, ASP Conference Series, {\it The
Fundamental Properties of Dwarf Elliptical Galaxies in Clusters}
\bibitem[]{} Held, E. V., \& Mould, J. R., 1994, AJ, 107, 1307
\bibitem[]{} Held, E. V., de Zeeuw, T., Mould, J., \& Picard, A., 1992, AJ, 103,851
\bibitem[]{} Hudson, M. J., Lucey, J. R., Smith, R. J., Steel, J., 1997, MNRAS, 291, 488
\bibitem[]{} Hudson, M. J., Lucey, J. R., Smith, R. J., Schlegel,
D. J., \& Davies, R. L., 2001, MNRAS, 327, 265
\bibitem[]{} J$\o$rgensen, I., Franx, M., \& Kj$\ae$rgaard, P., 1995, MNRAS, 276, 1341
\bibitem[]{} Jerjen, H, \& Binggeli, B., 1997, in ASP Conf. Ser. 116,
The Nature of Elliptical Galaxies, ed. M. Arnaboldi, G. S. Da Costa,
\& P. Saha (San Francisco: ASP), 239
\bibitem[]{} Jerjen, H, Binggeli, B., \&Freeman, K. C., 2000a, AJ,
119, 593
\bibitem[]{} Karachentsev, I.D., Karachentseva, V.E., Richter, G.M.,
\& Vennin, J.A., 1995, Astronomy and Astrophysics, 296,643
\bibitem[]{} Kormendy, J., 1985, ApJ, 295, 73
\bibitem[]{} Mehlert, D., Noll, S., Appenzeller, I., Saglia, R. P.,
Bender, R., B\"ohm, A., Drory, N., Fricke, K., Gabasch, A., Heidt, J.,
and 6 coauthors, 2002, A\&A, 393, 809
\bibitem[]{} Mobasher, B., Bridges, T. J., Carter, D., Poggianti,
B. M., Komiyama, Y., Kashikawa, N., Doi, M., Iye, M., Okamura, S.,
Sekiguchi, M., Shimasaku, K., Yagi, M., \& Yasuda, N., 2002, ApJS,
137, 279
\bibitem[]{} Moore, B., Lake, G., \& Katz, N., 1998, ApJ, 495, 139
\bibitem[]{} Moore, B., Lake, G., \& Katz, N., 1996, IAU Circ., 171, 203
\bibitem[]{} Moore, S. A., Lucey, J. R., Kuntschner, H., \& Colless,
M., 2002, MNRAS, 336, 382
\bibitem[]{} Pedraz, S., Gorgas, J., Cariel, N., Sanchez-Blazquez, P.,
Guzm\'an, R., 2002, MNRAS, 332, L59
\bibitem[]{} Peterson, R. C., \& Caldwell, N., 1993, AJ, 105, 1411
\bibitem[]{} Peterson, R. C., \& Caldwell, N., 1983, AJ, 105, 1411
\bibitem[]{} Prugniel, Ph, Bica, E., Klotz, A., \& Alloin, D., 1993,
A\&AS, 98, 229
\bibitem[]{} Sandage, A., \& Binggeli, B., 1984, AJ, 89, 919
\bibitem[]{} Sargent, W. L. W., Schechter, P. L., Boksenberg, A., \&
Shortridge, K., 1977, ApJ, 212, 326
\bibitem[]{} Schechter, P. L., \& Gunn, J. E., 1979, ApJ, 229, 472
\bibitem[]{} Schechter, P. L., 1980, AJ, 85, 801
\bibitem[]{} Simien, F., \& Prugniel, P., 2002, A\&A, 384, 371
\bibitem[]{} Smith, R.J., Hudson, M.J., Nelan, J.E., Moore, S.A.,
Quinney, S.J., and 6 co-authors, 2004, AJ, 128, 1558S
\bibitem[]{} Terlevich, A. I., Davies, R. L., Faber, S. M., \&
Burstein, D., 1981, MNRAS, 196, 381
\bibitem[]{} Terlevich, A. I., Caldwell, N., \& Bower, R. G., 2001,
MNRAS, 326, 154
\bibitem[]{} Tonry, J.L., 1981, ApJ, 251, 1
\bibitem[]{} Tonry, J.L., \& Davis, M., 1981, ApJ, 246, 680
\bibitem[]{} van Zee, L., Skillman, E.D., \& Haynes, M.P., 2004, AJ, 128, 121
\bibitem[]{} Wirth, A., \& Gallagher, J. S., 1984, ApJ, 282, 85
\end{thebibliography}
\end{document}